\documentclass[11pt]{article}
\usepackage{moriond,epsfig}

\def\Journal#1#2#3#4{{#1} {\bf #2}, #3 (#4)}

\def\be{\begin{equation}}
\def\ee{\end{equation}}
\def\bea{\begin{eqnarray}}
\def\eea{\end{eqnarray}}

\def\Msun{$\rm M_\odot$}
\def\Lsun{$\rm L_\odot$}
\def\Teff{$T_{\rm eff}$}

\def\simgt{\lower.5ex\hbox{$\; \buildrel > \over \sim \;$}}
\def\simlt{\lower.5ex\hbox{$\; \buildrel < \over \sim \;$}}
\def\ltsima{$\; \buildrel < \over \sim \;$}
\def\gtsima{$\; \buildrel > \over \sim \;$} 
\def\lsim{\lower.5ex\hbox{\ltsima}} 
\def\gsim{\lower.5ex\hbox{\gtsima}} 
 
\def\psr{PSR~J1740-5340}
\def\pspin{P_{\rm spin}}
\def\pdot{\dot{P}_{\rm spin}}

\begin{document}
\vspace*{4cm}
\title{Radio ejection in the evolution of X-ray binaries: the bridge between
low mass X-ray binaries and millisecond pulsars}

\author{Luciano Burderi, Francesca D'Antona}

\address{Osservatorio Astronomico di Roma, Via Frascati 33, 
             00040 Monteporzio Catone (Roma), Italy}

\author{Tiziana Di Salvo}

\address{Astronomical Institute "Anton Pannekoek," University of 
              Amsterdam, Kruislaan 403, NL 1098 SJ Amsterdam, 
              the Netherlands}

\author{Marta Burgay}

\address{Universit\`a di Bologna, Dipartimento di Astronomia, 
               via Ranzani 1, 40127 Bologna, Italy}

\maketitle\abstracts{We present a scenario for the spin-up 
and evolution of binary millisecond pulsars.  This can explain the 
observational properties of the recently discovered binary millisecond 
pulsar \psr, with orbital  period 32.5 hrs, in the Globular Cluster  NGC 6397.
The optical counterpart of this system is a star  as luminous as the cluster 
turnoff stars,  but with a lower \Teff\ (a  larger radius) which we  model 
with a star  of initial mass compatible  with the  masses evolving  in the  
cluster  ($\simeq 0.85$ \Msun). This star has  suffered Roche lobe overflow 
while  evolving off the main sequence, spinning up the neutron star to the 
present period  of 3.65 ms.  There are evidences that at  present, Roche lobe 
overflow is still going on. Indeed Roche lobe deformation of the mass losing
component is necessary to  be compatible with the  optical light curve.  
The presence of matter around the system is also consistent with the long 
lasting irregular radio eclipses seen in the system. 
We propose that this system is presently in a phase of `radio--ejection' 
mass loss. The radio--ejection phase can be initiated only if the system is 
subject to intermittency in the mass  transfer during the spin--up phase.
In fact, when the system is detached the pulsar radio emission is
not quenched, and may be able to prevent further mass accretion due to
the action  of the pulsar pressure  at the inner  Lagrangian point.
}

\section{Introduction}

The widely accepted scenario for the formation of a millisecond radio pulsar
(hereafter MSP) is the  recycling of an old  neutron star (hereafter NS)  
by a spin-up process  driven by  accretion of  matter and  angular momentum  
from a Keplerian disc, fueled {\it via}  Roche lobe overflow of a binary 
late--type companion (see  Bhattacharya  \&  van den  Heuvel  1991 for  a
review). If  the NS has a  magnetic dipole moment  (typical values are
$\mu \sim 10^{26}\--10^{27}$ G cm$^{3}$)  the disc is truncated at the
magnetosphere,  where   the  disc  pressure  is balanced by
the magnetic pressure, $P_{\rm MAG}$, exerted by the NS magnetic  field.   
Once  the accretion  and spin-up  process
ends, the NS is visible as a MSP.
Indeed,  a  common requirement  of  all  the  models of  the  emission
mechanism  from   a  rotating  magnetic  dipole  is   that  the  space
surrounding the NS  is free of matter up to  the light cylinder radius
$R_{\rm LC}$ (at  which the speed of a body  rigidly rotating with the
NS equals the speed of light) .

An interesting evolutionary phase  can occur if the mass transfer  
rate drops below the  level required to allow the expansion of the 
magnetosphere beyond $R_{\rm  LC}$, switching--on  
the emission from the rotating magnetic dipole (e.g. Illarionov \&
Sunyaev, 1975; Ruderman, Shaham \& Tavani 1989; Stella et al., 1994).
The pressure exerted by the radiation field of the radio pulsar
may overcome the pressure of the accretion disk, thus determining the 
ejection of matter from the system. Once the disk has been swept away, 
the radiation pressure stops the infalling matter as it overflows the 
inner Lagrangian point.

\section{The Effects of the Pulsar Energy Output}

The push on the accretion flow exerted by the (assumed dipolar) magnetic 
field of the NS can be described in terms of an outward pressure (we use the 
expressions outward or inward pressures to indicate the direction of the 
force with respect to the radial direction) exerted by the (assumed dipolar) 
rotating magnetic field, $B$, on the accretion flow: 
$P_{\rm MAG} = \frac{B^2}{8\pi} = 7.96 \times 10^{14} \mu_{26}^2 r_6^{-6}
\;\; {\rm dy}/{\rm cm}^2$, where $\mu_{26}$ is the magnetic moment of the NS 
in units of $10^{26}$ G cm$^3$, and $r_6$ is the distance from the NS center 
in units of $10^6$ cm.
Another contribution to the outward pressure is given by the radiation 
pressure (if present) generated by the rotating magnetic dipole, which, 
assuming isotropic emission, is given by:
$P_{\rm RAD} = 2.04 \times 10^{12} P_{-3}^{-4} 
\mu_{26}^2 r_6^{-2} \;\; {\rm dy}/{\rm cm}^2$.
The accretion flow, in turns, exerts a inward pressure on the field: 
for a Sakura--Sunyaev accretion disc in zone C (certainly the case at large
radii), the pressure is dominated by the gas contribution, 
$P_{\rm DISC} \propto \dot{M}^{17/20} r^{-21/8}$, where 
$\dot{M}$ is the accretion rate (see e.g.\ Frank, King \& Raine 1992).
This inward ram pressure of the accretion flow equals the outward pressure 
due to the magnetic dipole at the magnetospheric radius that defines an 
equilibrium point.  
Since the  radial dependence of the magnetostatic pressure exerted by the 
NS is steeper than the radial dependence of the disc pressure, the equilibrium
point defined above is stable inside $R_{\rm LC}$, where the 
pulsar radiation pressure $P_{\rm RAD}$ is absent. On the other hand,
the equilibrium point is unstable beyond $R_{\rm LC}$.
Therefore if $P_{\rm RAD}$ dominates over $P_{\rm DISC}$ for  any $r > R_{\rm
LC}$~ the whole accretion  disc is swept away  up to the  inner
Lagrangian point $L_1$ by  the radiation  pressure of the  pulsar. During
this ``radio ejection'' phase,  the mechanism that drives {\it  mass
overflow} from $L_1$ can  well be active, but  {\it the pulsar  radiation
pressure at $L_1$ prevents mass accretion onto the NS} (see Burderi et al. 
2001).  

It is possible that mass transfer during the binary system evolution 
suffers instabilities, due to the  X-ray illumination  of  the secondary  
star during the accretion phases, which makes the mass transfer rate quite 
unsteady.  In this case the system temporarily 
detaches, allowing  the pulsar to switch--on. However the companion 
evolution, which leads  to radius  expansion, will lead again to overflow. 
Mass transfer to the NS  and spin--up, due to the accretion of angular 
momentum of the  mass overflowing  the  Roche  lobe can therefore go  on, 
until  the pulsar  has been so much spun up that its radiation pressure 
at the inner Lagrangian point  is high  enough  to prevent mass  accretion. 
In  this case  we  expect to  be in  the presence of  a radio MSP  which 
from time  to time is obscured  by the matter floating around the system.

\section{\psr}

The  eclipsing MSP  \psr, discovered by  D'Amico et al. (2001a)
in the  globular cluster  NGC 6397, has the longest orbital
period ($P_{\rm  orb} \simeq 32.5$  hrs) and the most  massive minimum
companion mass (0.18 \Msun) among the 10 eclipsing pulsars detected up
to now.   The spin period ($\pspin  \simeq 3.65 \times 10^{-3}$  s) and its
derivative  ($\pdot =  1.59 \times  10^{-19}$), recently  derived by
D'Amico et  al. (2001b), allow  the determination of the  NS magnetic
moment, $\mu_{26} \simeq 7.7$.  Its position with respect to the cluster 
center excludes the possibility of a  contamination of $\pdot$ due to the NS
acceleration in  the gravitational field of the  cluster 
(D'Amico et al. 2001b), implying  that the estimate of the NS magnetic
moment is  reliable.
The optical counterpart of \psr, identified by Ferraro et al. 2001 with a 
slightly evolved turnoff star  in  the sample  studied  by Taylor  et  al. 
(2001) using data from the HST archive, also shows light modulation at the 
same orbital period as the radio data (Ferraro et al. 2001).  

\psr\ shows radio eclipses lasting for  about 40\% of the orbital phase 
at 1.4 GHz (D'Amico et al. 2001b).  Out of eclipse  the pulsar signal at  1.4
GHz shows significant  excess propagation delays  (up to $\sim 3$  ms) and
strong intensity  variations.  
This  suggest that in \psr\ the  signal is propagating
through a dense  material surrounding the system.  In order to  investigate
this possibility, D'Amico et al. (2001b) have fitted  the excess delays
measured in  two adjacent bands of 128 MHz each at 1.4  GHz. They found that
the excess delays $\Delta t$  can   be  well  fitted   with  the  equation
$\Delta   t  \propto \nu^{-2.02\pm0.30}$  that strongly  supports the
hypothesis  that the responsible mechanism is dispersion in a ionized medium
(see Fig.~\ref{fig:fig1}, right panel).
In this case the  corresponding  electron  column  density variations  are
$\Delta n_{\rm  e}  \sim 8  \times  10^{17}  \Delta  t_{-3}$ cm$^{-2}$,
where $\Delta t_{-3}$ is the delay at 1.4 GHz in ms. For $\Delta t_{-3} \sim
3$ the estimated electron column  density is $\sim 2.4 \times 10^{18}$
cm$^{-2}$.

The  eclipsing radius of the system is $R_{\rm E}  \sim 4.4 \times  
10^{11}$ cm (D'Amico et  al. 2001b), taking
$m_1=1.8$\Msun\ for the NS mass and $m_2=0.45$\Msun\ for the secondary mass
(see below) and $P_{\rm  orb} \sim 32.5$ hr.
This radius is  larger than the Roche lobe radius of the  secondary ($\sim
1.3 \times 10^{11}$ cm). This means that the eclipsing  matter is beyond the
gravitational influence of the  companion star and  must be continuously
replenished. From a simple calculation (see Burderi et al. 2002 for details), 
we can estimate a rough order of magnitude of the necessary
mass loss rate from the secondary, $\dot{M}$, by assuming spherical symmetry 
(which is, however, not consistent with the randomly variable signal intensity 
shown by the radio data). We find $\dot{M}$ up to $\sim 0.6 \times 10^{-10}$ 
\Msun\ yr$^{-1}$. Even considering the uncertainty on this estimate,
winds induced by the pulsar radiation are typically 2--3 orders of
magnitude weaker (Tavani \& Brookshaw, 1991). It is also unlikely that 
this matter is provided by the wind from the main sequence companion star,
given that the mass loss rate due to the star wind is expected to be less 
than $\sim 10^{-12}$ \Msun\ yr$^{-1}$.
We conclude that the  mass loss rates we derive are more consistent with
Roche lobe overflow driven by nuclear evolution of the secondary and orbital
angular momentum mass loss, than with a possible wind from the secondary.

We propose that this system is now experiencing the radio  ejection phase 
postulated above.  When: i) as a result
of the accretion of matter and angular momentum, the NS spin period is
so short that,  potentially, the radiation pressure of  a pulsar phase
would  be  high  enough  to   overcome  the  pressure  of  the  matter
overflowing the Roche  lobe, and ii) the  oscillations in $\dot{M}$
are large enough to allow the MSP to switch--on, then the
radio--ejection phase begins, leading  to the appearance of the system
as it looks now.

\section{Binary Evolution in the Globular Cluster NGC 6397} 

\begin{figure}
\psfig{figure=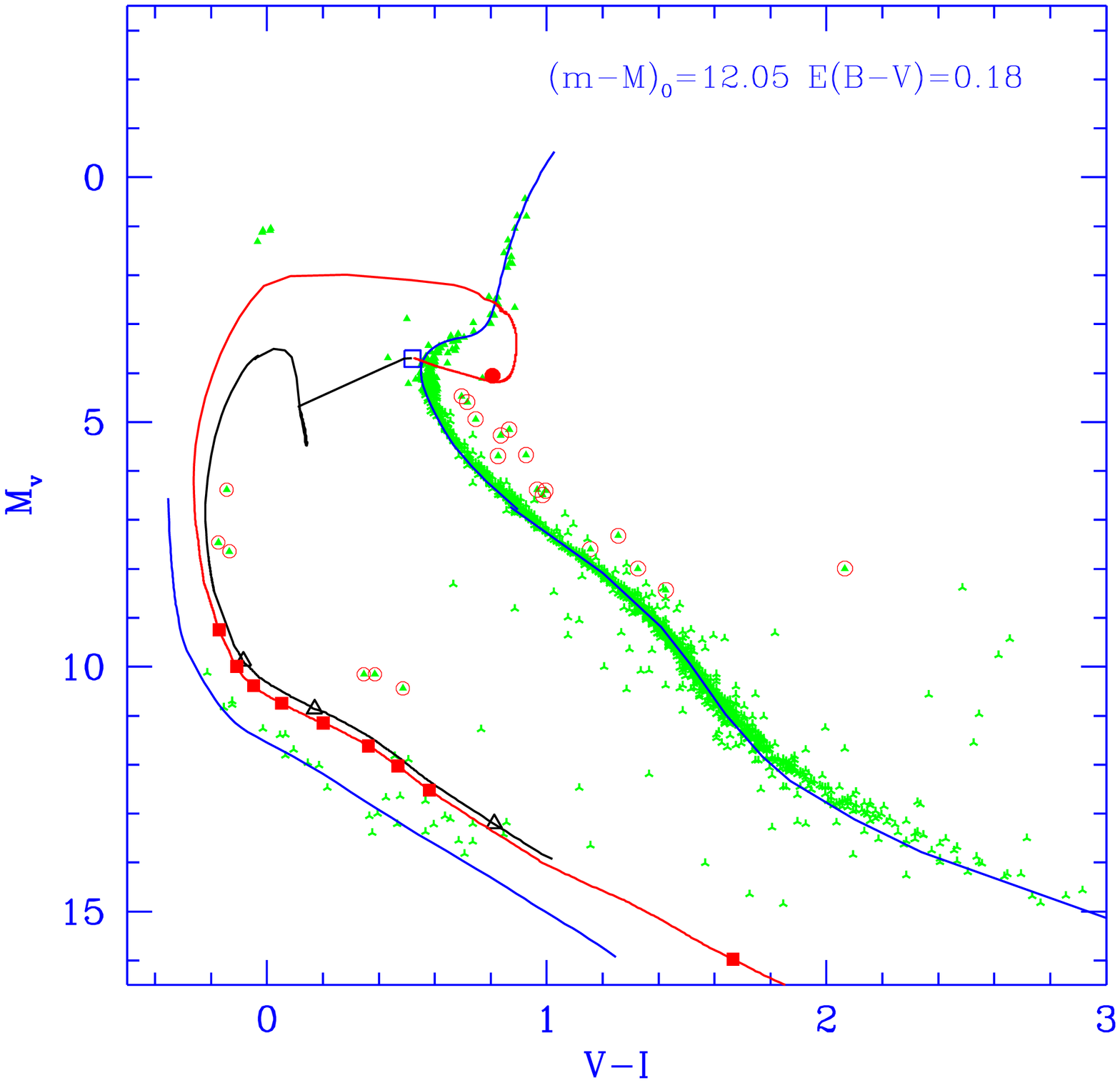,height=3.2in}
\psfig{figure=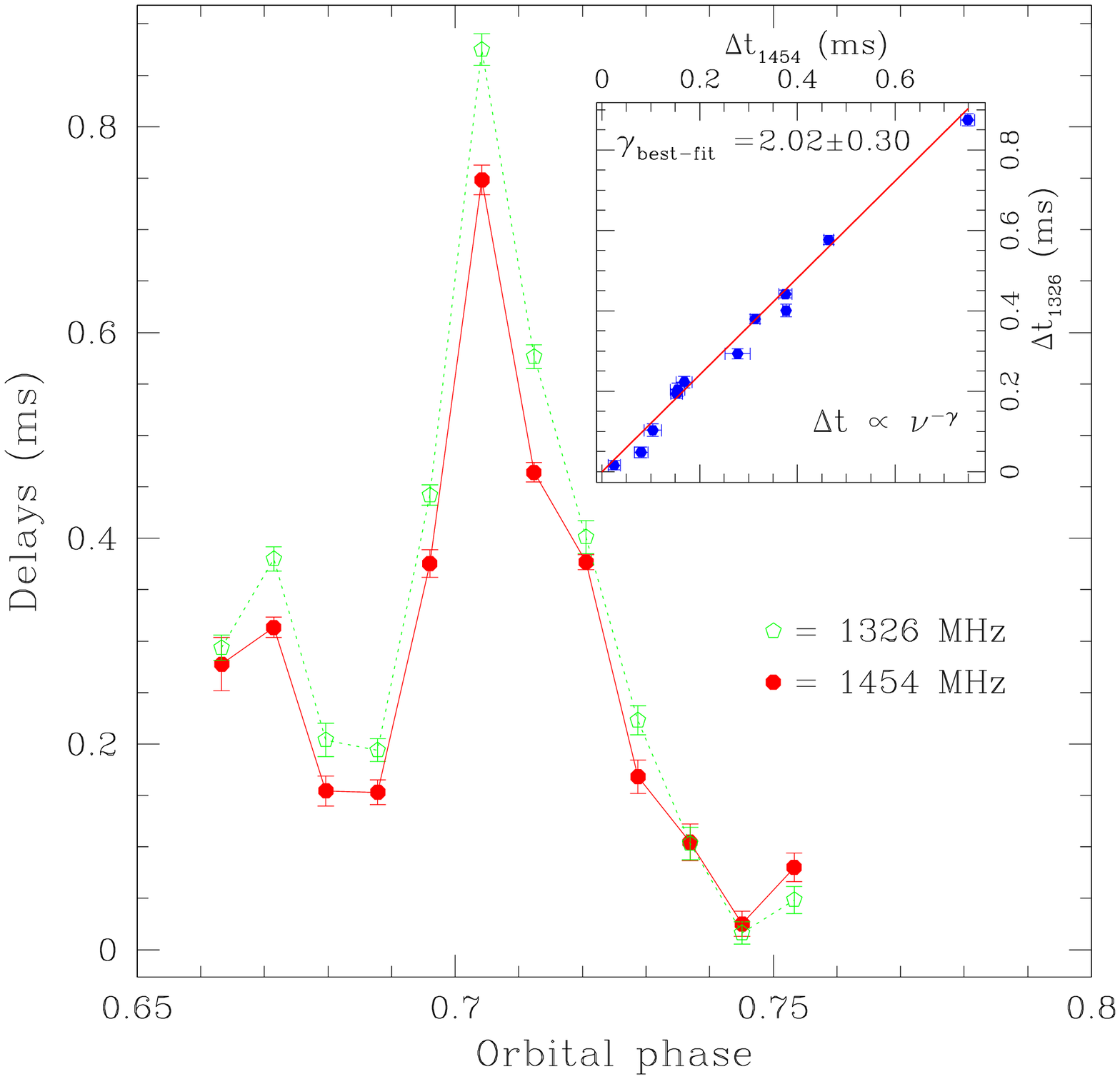,height=3.2in}
\caption{{\bf Left:} HR diagram of the Globular Cluster NGC 6397, with
an  isochrone of 12 Gyr superimposed. Starting close  to
the  turnoff at  the  open  square,  we show a standard evolution (case
2) of the companion  of the NS, having an initial mass of  0.85\Msun\ and  an
initial  orbital period  of  14.4 hrs (track evolving first  towards the
right  of the figure, and  passing through the observational counterpart of
\psr\  -- full dot), and the evolution of case 3, which includes a fixed
irradiation luminosity (track to the left).   Both sequences end  into the
WD evolution  of a  0.246 \Msun\ (case 2) and 0.216 \Msun\ (case 3). Both
sequences are followed along the WD cooling,  which  is dominated  by
residual proton  proton burning lasting for more than a Hubble time. Ages of
17, 20 and 26 Gyr are labelled as triangles along the cooling track of case 3.
Ages from 13 to 20 in steps of 1 Gyr, plus a last point at 25 Gyr, are 
labelled as squares for case 2.
{\bf Right:} Excess propagation delays as a function of the orbital phase
in \psr. The inset shows the best fit to these time delays.} 
\label{fig:fig1} 
\end{figure}

In order  to choose coherently  the input parameters for  the possible
evolution leading to  \psr, it is important to  take into account what
we know of the general properties  of the host cluster.  Figure \ref{fig:fig1}
(left panel) shows the composite HR diagram of NGC 6397 in the plane
$M_v$ versus $V-I$. The  open circles identify the objects  examined by 
Taylor  et al. (2001) in the core of this  cluster, to select objects which 
have been probably subject to binary evolution.  One  of the  Taylor et  al. 
(2001)  BY Dra candidates, plotted  as a  full dot in  Figure~\ref{fig:fig1} 
(left panel), is indeed the  optical counterpart of \psr. 
On the observational HR diagram we show an isochrone of 12 Gyr for metallicity 
in mass  fraction Z=0.006 and helium mass fraction Y=0.23.  The isochrone of 
12 Gyr, which  fits the cluster  HR diagram, implies  that a mass  of $\sim
0.81$ \Msun\ is  evolving at the cluster turnoff (TO),  and that its TO
luminosity is  $\simeq 2.24$ \Lsun.  Different  interpretations of the
HR diagram  morphology, assuming that  the distance of the  cluster is
smaller,  may lead  to values  of the  TO luminosity  down  $\sim 1.8$
\Lsun.
The optical  counterpart of \psr\ is at a luminosity similar
to  the   TO  luminosity  (i.e.\ 1.8-2.3   \Lsun), but cooler than the TO,  
that is, at a radius larger than the TO radii. 

We study four cases of evolution (see Fig.~\ref{fig:fig1}, left panel, and 
Burderi et al. 2002). In all cases, 
we  model the initial parameter of the system as starting  with a 0.85 
\Msun\ companion, a 1.4\Msun\ NS, and a orbital initial period of 14.27 hr.
We  follow  the binary  evolution  with  the  ATON1.2 code  (D'Antona,
Mazzitelli, \& Ritter 1989). The  mass loss rate is computed following
the formulation  by Ritter (1988),  as an exponential function  of the
distance of  the stellar  radius to  the Roche lobe,  in units  of the
pressure scale  height. This method  also allows to compute  the first
phases of  mass transfer, during  which the rate reaches  values which
can  be much larger  than the  stationary values,  due to  the thermal
response of the star to mass  loss. The evolution of the system also
includes orbital angular momentum losses through magnetic braking, in the
Verbunt \& Zwaan (1981) formulation, in which the braking parameter is
set to $f=1$. We also tested a case in which $f=2$.

When mass accretion on the NS is assumed, and the binary evolution suffers a
Low Mass X-ray Binary (LMXB) phase, it is also important to consider how the 
X--ray phase would affect the binary evolution we are considering, as a 
fraction $R_2^2/(2a)^2 \le 2.7 \times 10^{-2}$ of the X--ray luminosity 
impacts at the secondary surface. 
A self--consistent modelling is quite difficult, as the feedbacks  are not 
easy  to describe both  physically and  numerically. We follow a simplified
schematization in which the star is immersed in the X--ray radiation bath, 
and the total luminosity $L_{tot}$\ which it must radiate is the sum of the 
stellar luminosity $L_*$\ plus the heating luminosity $L_h$, from which an
`irradiation temperature' $T_{irr}=(L_h/4 \pi \sigma R_2^2)^{1/2}$\ is defined.
The stellar luminosity and \Teff\ then are related by:
$L= L_{tot}-L_h=4 \pi \sigma R_2^2(T_{\rm eff}^4-T_{irr}^4)$
Consequently, the stellar \Teff\  becomes hotter due to  the irradiation, and
the star rapidly evolves in the HR diagram at a location determined by the
amount of  irradiation allowed.  The  phases  of  mass transfer  are only
slightly altered by  the new system conditions. 
However, as the star loses mass, its radius becomes larger
than the  radius  of the  standard  sequence. This  difference amounts to
$\sim 20$\%  at the orbital  period of \psr. 

All the sequences are evolved until the mass loss phase finally ends with the
stellar remnant evolving into the white dwarf region as low mass helium white
dwarfs. We see that the most luminous three  objects among those identified by
Taylor et al. (2001) as helium WDs  actually may be the end--products of such
an evolution. 
The optical component of \psr\ also seems compatible with the evolution we
have suggested. It lies along case 2 evolution, that not including
irradiation, as, in fact, the pulsar luminosity is not important as
irradiation source for the binary.

\subsection{The Evolution of the Progenitor System of \psr}
As already said, the binary evolution which we are
describing is not dramatically altered by the irradiation due to the LMXB
phase: in particular, the final evolution is very similar, although the
white dwarf remnant mass is slightly smaller. However, it helps to predict
that the LMXB phases can be alternated with phases in which the system
remains detached, as already had been suggested for systems having main 
sequence companions. We now consider the evolution of the system, taking 
also into account the pulsar's behavior.
Any time the system detaches, the radio pulsar will switch on, but it
will be  quenched  again  when  the  mass transfer  resumes. In the HR
diagram the position of the secondary component will shift from its
`irradiated' position during the LMXB mass transfer phase (track 3 on the
left of the MS) to its `standard' position along track 1. However,  as
discussed  in Burderi  et al. (2001), when the pulsar is spinning
sufficiently  fast  and  the   orbital  period  is  sufficiently  long
(longer than a critical value $P_{\rm crit} \propto \pspin^{9.6}~ 
\mu^{-4.8}~ \dot{M}^{2.04}$), the radiation pressure 
exerted by the pulsar  at the inner inner Lagrangian point  is larger than 
the pressure exerted by the matter  overflowing the Roche lobe even if the
mass transfer rate recovers its  secular value dictated by the nuclear
evolution of  the companion.
If  $P_{\rm orb} \ge P_{\rm crit}$ {\it
the system  will remain  in the radio--ejection  phase during  all the
subsequent binary evolution}. In this case, the matter overflowing the Roche
lobe will  be accelerated  by its  interaction with  the pulsar radiation and
ejected from the system. 
In the case of \psr, the critical period $P_{\rm crit}$\ to reach
the `radio--ejection' phase is $\sim 39$ hr, not very different from its
orbital period.  Thus, \psr\ is possibly in the radio--ejection phase. 
Considering the dependency of $P_{\rm crit}$ from the system parameters, 
the fact that $P_{\rm crit}  \sim  P_{\rm orb}$  is compelling.

\section{Summary and Conclusions} 
\label{sec:system}
\begin{figure}
\centering
\psfig{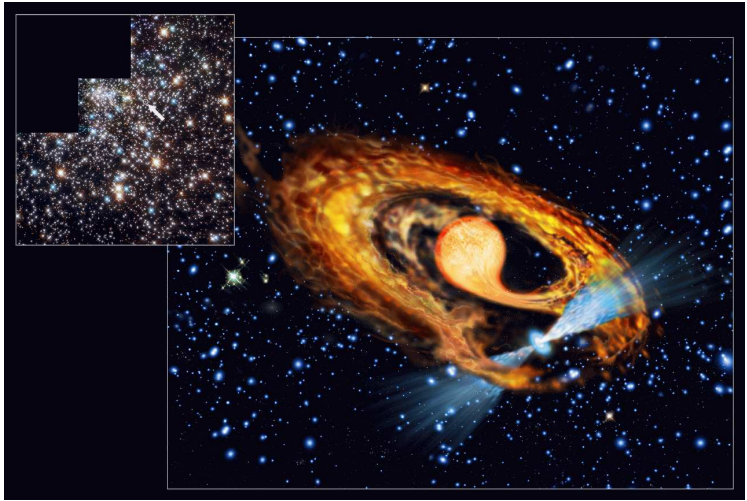}
\caption{An artistic impression of the system \psr\ based on the scenario
proposed in this paper (courtesy of ESO).
\label{artistic}}
\end{figure}

We have considered the evolution of possible progenitors of the binary MSP
\psr\ in the Globular Cluster NGC 6397.
We can reproduce the HR diagram location of the optical companion, starting
mass transfer to the NS from a hypothetical secondary of mass 0.85 \Msun,
slightly evolved off the main sequence when mass transfer begins.
In conclusion we propose that:

i) Orbital evolution  calculations shows that a slightly
evolved 0.85 \Msun\ secondary orbiting a NS can transfer mass to the NS, and
reaches a stage in which its mass is reduced to $\simeq 0.45$ \Msun, and its
optical location in the HR diagram is then
compatible with the recently detected optical counterpart of \psr;

ii) \psr\ might represent a system whose evolution has been
envisioned by Burderi et al. (2001): the spin and the magnetic moment of the
pulsar may keep the system in a radio--ejection  phase in which accretion is
inhibited by the  radiation pressure  exerted by the  pulsar on  the
overflowing matter while  the mechanism that  drives the Roche lobe  overflow
from the companion is still active, thus causing an intense wind which
would be very difficult to explain otherwise.
This evolution seems to be the only viable possibility to explain the long
lasting eclipses and the strong intensity variation randomly occurring
in the radio emission.  An artistic impression of the system, according to
this scenario, is shown in Figure~\ref{artistic}.

As a final remark we note that $P_{\rm orb} \sim P_{\rm crit}$ suggests
the interesting possibility that this system could swiftly switch from
the  present radio  pulsar phase  to an  accretion phase  in  which it
should be visible as a $L_{\rm X} \sim 10^{36}$ ergs s$^{-1}$ LMXB.

\section*{Acknowledgments}
This  work was  partially supported  by a grant   from  the   Italian  
Ministry   of  University   and  Research (Cofin-99-02-02).

\end{document}